\documentclass[pra,twocolumn,floatfix,a4paper,superscriptaddress]{revtex4}

\usepackage{bm,color,amsmath,txfonts}
\usepackage{graphicx}
\usepackage{siunitx}
\usepackage{subfigure}
\usepackage{verbatim}
\usepackage{dcolumn}
\usepackage{bm}
\usepackage{epsf}
\usepackage{xcolor}
\usepackage{hyperref}
\usepackage{hhline}
\usepackage{braket}
\usepackage{float}
\usepackage{enumerate}
\usepackage{bbm}
\usepackage{lipsum}
\usepackage{appendix}

\begin{document}

	\title{Ground-state cooling of a massive mechanical oscillator by feedback in cavity magnomechanics}	
	\author{Zhi-Yuan Fan}
	\affiliation{Interdisciplinary Center of Quantum Information, State Key Laboratory of Modern Optical Instrumentation, and Zhejiang Province Key Laboratory of Quantum Technology and Device, School of Physics, Zhejiang University, Hangzhou 310027, China}
	\author{Hang Qian}
	\affiliation{Interdisciplinary Center of Quantum Information, State Key Laboratory of Modern Optical Instrumentation, and Zhejiang Province Key Laboratory of Quantum Technology and Device, School of Physics, Zhejiang University, Hangzhou 310027, China}
	\author{Xuan Zuo}
	\affiliation{Interdisciplinary Center of Quantum Information, State Key Laboratory of Modern Optical Instrumentation, and Zhejiang Province Key Laboratory of Quantum Technology and Device, School of Physics, Zhejiang University, Hangzhou 310027, China}
	\author{Jie Li}\thanks{jieli007@zju.edu.cn}
	\affiliation{Interdisciplinary Center of Quantum Information, State Key Laboratory of Modern Optical Instrumentation, and Zhejiang Province Key Laboratory of Quantum Technology and Device, School of Physics, Zhejiang University, Hangzhou 310027, China}

\begin{abstract}
Cooling the motion of a massive mechanical oscillator into its quantum ground state plays an essential role in observing macroscopic quantum effects in mechanical systems. Here we propose a measurement-based feedback cooling protocol in cavity magnomechanics that is able to cool the mechanical vibration mode of a macroscopic ferromagnet into its ground state.   The mechanical mode couples to a magnon mode via a dispersive magnetostrictive interaction, and the latter further couples to a microwave cavity mode via the magnetic-dipole interaction. A feedback loop is introduced by measuring the amplitude of the microwave cavity output field and applying a force onto the mechanical oscillator that is proportional to the amplitude fluctuation of the output field. We show that by properly designing the feedback gain, the mechanical damping rate can be significantly enhanced while the mechanical frequency remains unaffected.  Consequently, the vibration mode can be cooled into its quantum ground state in the unresolved-sideband regime at cryogenic temperatures. The protocol is designed for cavity magnomechanical systems using ferromagnetic materials which possess strong magnetostriction along with large magnon dissipation.
\end{abstract}
	\date{\today}
	\maketitle

\section{Introduction}
	
Cavity magnomechanics (CMM) is a newly developed system that studies coherent interactions between microwave cavity photons, collective spin excitations (magnons) and magnetostriction-induced vibration phonons in magnetically ordered materials~\cite{Tang,Jie18,Jie20,Davis,Li22}.  In typical CMM experiments~\cite{Tang,Davis,Li22}, a macroscopic ferrimagnetic yttrium-iron-garnet (YIG) sphere with the diameter in the $10^2$ $\mu$m range is adopted, which supports a gigahertz magnon mode and a megahertz mechanical vibration mode. The much lower mechanical frequency permits a dominant magnon-phonon dispersive coupling~\cite{QST23}. Such a radiation-pressure-like magnomechanical coupling enables the mechanical mode to be significantly cooled by driving the magnon mode with a red-detuned microwave field~\cite{Jie18,Jie20,Ding20}, or alternatively, by driving the cavity polariton with a red-detuned pump field in the cavity-magnon strong coupling regime~\cite{Davis,Li22}. The cooling mechanism is akin to the sideband cooling in cavity optomechanics~\cite{c1,c2,c3}. Cooling the low-frequency mechanical mode (close) to the ground state is of fundamental importance, because it is a prerequisite for preparing quantum states in the CMM system, e.g., entangled states~\cite{Jie18,Jie20,QST23,Jie19a,Tan,QST,Wu,Irfan,Nie,Ding22} and squeezed states~\cite{Jie19b,HFW,NSR}. The CMM system is gradually becoming a new platform for the study of macroscopic quantum states~\cite{rev,rev2}. Besides, it finds various potential applications in quantum information processing and quantum technologies~\cite{q1,q2,q3,q4,q5,q6,q7}.

To date, almost all the theoretical proposals and experiments in the field of CMM employ a YIG ferrimagnet, of which the magnon mode has a low dissipation rate, typically $\kappa_m/2\pi \sim 1$ MHz. This brings the system well within the resolved-sideband regime, where the magnon linewidth is much smaller than the mechanical frequency $\kappa_m \ll \omega_b$. In this regime, the sideband cooling is efficient and able to cool the mechanical motion into its ground state~\cite{Jie18,Jie20,Ding20,QST,Jie19b}. However, the drawback of the CMM based on YIG is that the bare magnomechanical coupling strength $g_m$ is rather weak, and $g_m \,{<}\, 10$ mHz for YIG spheres used in Refs.~\cite{Tang,Davis,Li22}. This means that a sufficiently strong drive power must be used to obtain a considerable magnomechanical effective coupling and cooperativity. A strong drive power, however, can bring about complex and complicated nonlinear effects~\cite{Li22}. Hence, the coupling enhancement by increasing the power is actually restricted, which places a limit on the effective mean phonon number that can be achieved by sideband cooling, and the degree of entanglement and squeezing that can be obtained in the protocols~\cite{Jie18,Jie20,Jie19a,Tan,QST,Jie19b,HFW,NSR}. Therefore, improving the bare magnomechanical coupling strength is of vital importance to all the studies related to CMM. To this end, one can simply reduce the size of the YIG sphere~\cite{Tang}. This, however, hinders us to observe quantum effects on a more macroscopic scale and, moreover, unwanted nonlinearity may become appreciable~\cite{Jie18}. One may then consider using other ferromagnetic materials~\cite{prb}, which exhibit much stronger magnetostriction, but also much larger magnon dissipation, thus bringing the system in the unresolved-sideband regime $\kappa_m > \omega_b$. The aforementioned sideband-cooling mechanism then becomes inefficient. Therefore, it is important to realize mechanical ground-state cooling in such materials with both strong magnetostriction and large magnon dissipation. A strategy has recently been offered, which shows that magnon squeezing can be exploited to efficiently cool the mechanical motion into its ground state in such unresolved-sideband regime~\cite{Asjad}. 

Here we provide a promising new approach to achieving mechanical ground-state cooling in the CMM systems with unresolved sidebands by introducing a measurement-based feedback loop. Specifically, we apply a feedback force onto the mechanical resonator that is proportional to the amplitude fluctuation of the cavity output field, which is measured by a homodyne setup. We show that by appropriately modulating the feedback gain, the mechanical mode can be significantly cooled in the unresolved-sideband regime, and can be cooled into its quantum ground state at cryogenic temperatures. We also analyse the effects of various experimental imperfections, such as nonunity detection efficiency and measurement noise, on the cooling results.  
	
The paper is organized as follows. In Sec.~\ref{syst}, we introduce the typical CMM system and provide its Hamiltonian and quantum Langevin equations (QLEs). In Sec.~\ref{fbCMM}, we show how to add a feedback loop to the CMM system, which consists of a microwave homodyne detection, a gain control module, and a feedback drive which applies the force onto the mechanical resonator. We derive the modified QLEs for such feedback-assisted CMM system and the mechanical noise spectrum density (NSD). We further analyse various noise sources heating the mechanical mode.  Finally, we present the numerical results in Sec.~\ref{result} and draw the conclusion in Sec.~\ref{conc}.

\section{The CMM system}\label{syst}
	
	\begin{figure}[h]
		\includegraphics[width=0.8\linewidth]{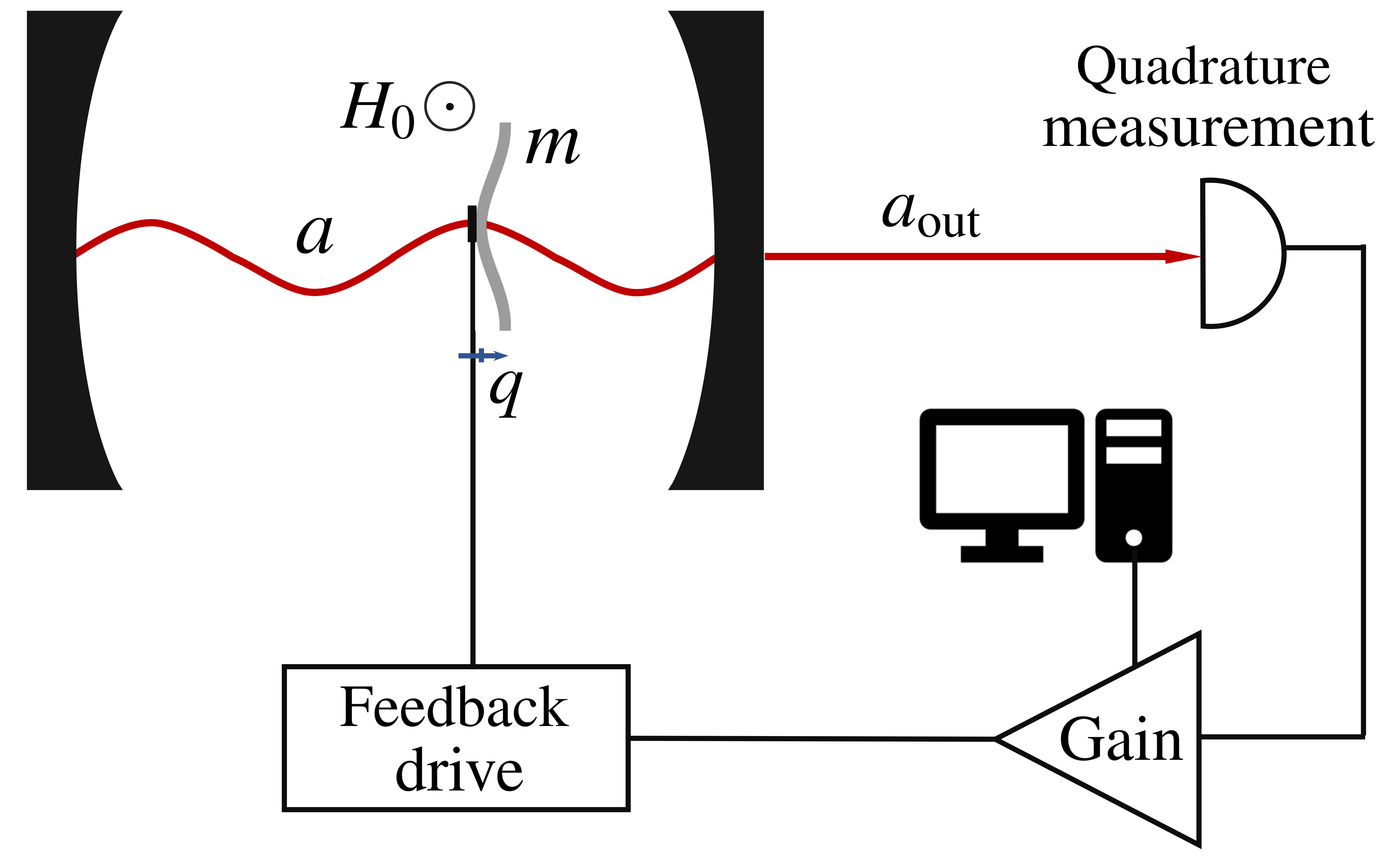}
		\caption{Sketch of the feedback-assisted CMM system. By measuring the amplitude of the microwave cavity output field, a gain module controlled feedback force is applied onto the mechanical resonator, which significantly cools the mechanical motion. }
		\label{fig1}
	\end{figure}
		
We start with the introduction of the typical CMM system~\cite{Tang,Davis,Li22}. It consists of a microwave cavity mode, a magnon mode, and a mechanical vibration mode; see Fig.~\ref{fig1}. A ferromagnet is placed inside a microwave cavity near the maximum magnetic field of the cavity mode and moreover in a uniform bias magnetic field, which activates the magnon mode (spin wave). Magnons are the quanta of collective spin excitations in magnetically ordered materials. Here, the magnon mode is of a particular ferromagnet that possesses strong magnetostriction but a large magnon dissipation rate, $\kappa_m > \omega_b$. This brings the magnomechanical system in the unresolved-sideband regime, where sideband cooling protocols become ineffective~\cite{c1,c2,c3}. The magnon mode couples to the deformation vibration mode of the ferromagnet via a dispersive magnetostrictive interaction, and to the microwave cavity mode via the magnetic-dipole interaction. The Hamiltonian of the CMM system reads
	\begin{align}
		\begin{split}
			H/\hbar=&\ \omega_a a^\dagger a+\omega_m m^\dagger m+\frac{\omega_b}{2}(q^2+p^2)+g_m m^\dagger mq\\
			&+g_a(am^\dagger +a^\dagger m)+i\Omega(m^\dagger e^{-i\omega_0 t}-m e^{i\omega_0 t}),
		\end{split}
	\end{align}
where $a$ ($a^\dagger$) and $m$ ($m^\dagger$) ($[j,j^\dagger]=1, j=a,m$) are the annihilation (creation) operators of the cavity mode and the magnon mode, respectively; $q$ and $p$ are the dimensionless position and momentum of the mechanical oscillator, satisfying $[q,p]=i$. $\omega_j$ ($j=a,m,b$) are the resonance frequencies of the cavity, magnon and mechanical modes, respectively. The magnon frequency can be adjusted by varying the bias magnetic field $H_0$ via $\omega_m=\gamma H_0$, with $\gamma$ being the gyromagnetic ratio. $g_m$ is the bare magnomechanical coupling strength and $g_a$ is the cavity-magnon coupling strength. Due to the cavity-magnon nearly resonant linear coupling, $g_a$ can be very strong, leading to cavity-magnon polaritons~\cite{Tang,Davis,Li22}, while the bare magnomechanical coupling $g_m$ is typically weak, but the effective magnomechanical coupling can be significantly enhanced by driving the magnon mode with a microwave field at frequency $\omega_0$.  The Rabi frequency $\Omega$ denotes the coupling strength between the drive magnetic field and the magnon mode, of which the expression depends on the specific material of the ferromagnet~\cite{Jie18}.

By including the dissipations and input noises into the system and working in the frame rotating at the drive frequency $\omega_0$, the dynamics of the system is governed by the following QLEs:
	\begin{align}\label{QLEs1}
		\begin{split}
			\dot{a}=&\ -(i\Delta_a+\kappa_a) a-ig_a m+\sqrt{2\kappa_a}a_{\mathrm{in}}, \\
			\dot{m}=&\ -(i\Delta_m+\kappa_m) m-ig_a a-i g_m mq+\Omega +\sqrt{2\kappa_m}m_{\mathrm{in}}, \\
			\dot{q}=&\ \omega_b p,\ \dot{p}=-\omega_b q-\gamma_b p-g_m m^\dagger m+\xi,
		\end{split}
	\end{align}
where $\Delta_{a(m)}=\omega_{a(m)}-\omega_0$, and $\kappa_a$, $\kappa_m$ and $\gamma_b$ are the dissipation rates of the cavity, magnon and mechanical modes, respectively. $a_{\mathrm{in}}$ and $m_{\mathrm{in}}$ denote the zero-mean input noises of the cavity and magnon modes, which are characterized by the correlation functions: $\langle j_{\mathrm{in}}(t) j_{\mathrm{in}}^\dagger(t') \rangle=\big[n_j(\omega_j)+1\big]\delta(t-t')$, $\langle j_{\mathrm{in}}^\dagger(t) j_{\mathrm{in}}(t') \rangle=n_j(\omega_j)\delta(t-t')$, $j=a,m$, where the equilibrium mean thermal excitation numbers $n_j(\omega_j)=1/[\mathrm{exp}(\hbar\omega_j/k_B T)-1]$, with $T$ being the bath temperature and $k_B$ as the Boltzmann constant.  $\xi$ is the Langevin force operator, which accounts for the Brownian motion of the mechanical resonator and is autocorrelated as~\cite{DV}
\begin{equation}
\langle \xi(t) \xi(t') \rangle = \frac{\gamma_b}{2\pi \omega_b} \int \omega e^{-i\omega(t-t')} \left[ \coth \left(\frac{\hbar \omega}{2k_B T} \right) +1  \right] {\rm d} \omega.
\end{equation}

	Under a strong and continuous microwave drive, the system evolves into a steady state at which the cavity and magnon modes have a large mean amplitude $\left|\langle a \rangle \right|, \left|\langle m \rangle \right|\gg 1$. This allows us to linearize the nonlinear dynamics around large mean values by writing each mode operator as a large classical average and a small quantum fluctuation, i.e., $O=\langle O \rangle+\delta O$, $O=a,m,p,q$, and by neglecting second-order fluctuation terms. Consequently, Eq.~\eqref{QLEs1} is separated into two sets of equations: one is for the classical averages and the other for the quantum fluctuations. By solving the former set of equations, we obtain the following steady-state solutions: 
	\begin{align}
		\begin{split}
			\langle m\rangle=&\ \frac{\Omega(\kappa_a+i\Delta_a)}{g_a^2+(\kappa_a+i\Delta_a)(\kappa_m+i\tilde{\Delta}_m)},\\
			\langle a\rangle=& \frac{-ig_a\Omega}{g_a^2+(\kappa_a+i\Delta_a)(\kappa_m+i\tilde{\Delta}_m)},  \\
			\langle p\rangle=& 0, \,\,\,\,\,\,  \langle q\rangle=-g_m\left|\langle m\rangle \right|^2/\omega_b,
		\end{split}
	\end{align}
	where $\tilde{\Delta}_m=\Delta_m+g_m \langle q\rangle$ is the effective magnon-drive detuning, including the magnon frequency shift caused by the magnomechanical interaction.

\section{Feedback-assisted CMM system}\label{fbCMM}

Here we introduce the central part of the protocol, i.e., the addition of a feedback loop to the conventional CMM system, as sketched in Fig.~\ref{fig1}. Specifically, a feedback force is applied onto the mechanical resonator (along the direction of the displacement $\vec q$~\cite{q6}), e.g., via a piezoelectric manner. The force is proportional to the fluctuation of the amplitude quadrature of the cavity output field measured via a homodyne setup~\cite{Davis}, which takes the form of  $ f_{\mathrm{fb}}(\omega) = -g(\omega) \delta X_a^{\mathrm{est}}(\omega)$, where $g(\omega)$ is the feedback gain function and can be realized by the gain module controlled via the Field Programmable Gate Array (FPGA)~\cite{Guo19FPGA}.  We consider the microwave drive field being resonant with the magnon mode and also with the cavity mode, i.e., $\Delta_a =\tilde{\Delta}_m = 0$. The resonant drive is adopted to implement measurement-based feedback cooling in cavity optomechanics~\cite{c3,98fbPRL,99fbPRL,02fbPRA,06fbNature,07fbPRL,15fbNature,18fbNature,21fbLJQ}, which is optimal for realizing high-sensitivity detection of the mechanical displacement. In this situation, the Stokes and anti-Stokes scattering probabilities are equal and the optomechanical interaction resembles a quantum nondemolition interaction~\cite{QND}.  However, unlike feedback cooling in the two-mode cavity optomechanics, where the phase quadrature of the cavity output field is measured, in the present three-mode CMM system, the magnon phase quadrature mediates the coupling between the mechanical displacement and the cavity amplitude quadrature. Therefore, in our system the amplitude of the microwave cavity output field should be measured, based on which a feedback force is applied.  

Consequently, we obtain the following QLEs for the quantum  fluctuations of the system quadratures in the frequency domain:
	\begin{widetext}
	\begin{subequations}
		\begin{align}
			-i\omega\delta X_a(\omega)=&-\kappa_a \delta X_a(\omega)+g_a\delta Y_m(\omega)+\sqrt{2\kappa_a}X^{\mathrm{in}}_a(\omega), \label{aaa} \\
			-i\omega\delta Y_a(\omega)=&-\kappa_a \delta Y_a(\omega)-g_a\delta X_m(\omega)+\sqrt{2\kappa_a}Y_a^{\mathrm{in}}(\omega),  \label{bbb} \\
			-i\omega\delta X_m(\omega)=&-\kappa_m \delta X_m(\omega)+g_a\delta Y_a(\omega)+\sqrt{2\kappa_m}X_m^{\mathrm{in}}(\omega), \label{ccc} \\
			-i\omega\delta Y_m(\omega)=&-\kappa_m \delta Y_m(\omega)-g_a\delta X_a(\omega)-G_m\delta q(\omega)+\sqrt{2\kappa_m}Y_m^{\mathrm{in}}(\omega), \label{ddd} \\
			-i\omega\delta q(\omega)=&\ \omega_b\delta p(\omega), \label{eee}  \\
			-i\omega \delta p(\omega)=&-\omega_b\delta q(\omega)-\gamma_b\delta p(\omega)-G_m \delta X_m(\omega)+\xi(\omega)+f_{\mathrm{fb}}(\omega),  \label{fff}
		\end{align}
	\end{subequations}
	\end{widetext}
where the quadrature fluctuations are defined as $\delta X_j=(\delta j+\delta j^\dagger)/\sqrt{2}$, $\delta Y_j=i(\delta j^\dagger-\delta j)/\sqrt{2}$, and $X_j^{\mathrm{in}}=(j_{\mathrm{in}}+j_{\mathrm{in}}^\dagger)/\sqrt{2}$, $\delta Y_j^{\mathrm{in}}=i(j_{\mathrm{in}}^\dagger-j_{\mathrm{in}})/\sqrt{2}$, $j=a,m$. $G_m=\sqrt{2}g_m\langle m\rangle$ is the effective magnomechanical coupling strength. Under the resonant condition $\Delta_a= \tilde{\Delta}_m=0$, the magnon amplitude $\langle m\rangle=\Omega \kappa_a/(g_a^2+\kappa_a\kappa_m)$, which leads to a real and positive coupling $G_m$. The last term $f_{\mathrm{fb}}(\omega)=-g(\omega)\delta X_a^{\mathrm{est}}(\omega)$ in Eq.~\eqref{fff} represents the feedback force that is applied onto the mechanical resonator. The measured (estimated) amplitude fluctuation of the microwave output field is
	\begin{equation}
\delta X_a^{\mathrm{est}}(\omega)= \frac{\sqrt{\eta}\ \delta X_a^{\mathrm{out}}(\omega) -\sqrt{1-\eta}\ X_\mathrm{vac}(\omega)}{\sqrt{2 \kappa_a} }+X_a^{\mathrm{imp}}(\omega),
	\end{equation}
where $\delta X_a^{\mathrm{out}}(\omega)= \sqrt{2\kappa_a}\delta X_a(\omega)-X_a^{\mathrm{in}}(\omega)$ is the amplitude fluctuation of the cavity output field, and the first term characterizes the detection of the cavity output field with detection efficiency $\eta \le 1$. The detection loss can be modeled by a beam splitter (with transmittance $\eta$ and reflectance $1-\eta$), where the reflection is treated as the detection loss~\cite{UL}. $X_\mathrm{vac}$ denotes the vacuum noise, and $X_a^{\mathrm{imp}}$ represents the measurement noise in the homodyne detection that leads to the imprecision of the amplitude quadrature~\cite{07fbPRL,15fbNature,18fbNature,RMPmeasure}, which is considered as white noise.

Solving Eqs.~\eqref{aaa} and \eqref{ddd} for the cavity amplitude fluctuation $\delta X_a(\omega)$, we obtain
	\begin{equation}\label{Xasolu}
		\begin{split}
		&\delta X_a(\omega)=\chi_a(\omega)\chi_{ma}(\omega)\ 
		\times\\
		&\left[-g_a G_m\delta q(\omega) + \chi_m(\omega)^{-1} \! \sqrt{2\kappa_a} X_a^{\mathrm{in}}(\omega) + g_a \sqrt{2\kappa_m} Y_m^{\mathrm{in}}(\omega)\right],
	\end{split}
	\end{equation}
 with $\chi_j(\omega)=1/(\kappa_j-i\omega)$, $j=a,m$, being the natural susceptibility of the cavity and magnon modes, respectively, and $\chi_{ma}(\omega)=[\chi_m(\omega)^{-1}+g_a^2\ \chi_a(\omega)]^{-1}$. Substituting Eq.~\eqref{Xasolu} into $\delta X_a^{\mathrm{est}}(\omega)$ and $f_{\mathrm{fb}}(\omega)$ in Eq.~\eqref{fff}, we extract the drift matrix $A$ from the QLEs Eqs.~\eqref{aaa}-\eqref{fff}, given by
	\begin{equation}\label{AAA}
		A=\begin{pmatrix}
			-\kappa_a & 0 & 0 & g_a & 0 & 0\\
			0 & -\kappa_a & -g_a & 0 & 0 & 0\\
			0 & g_a & -\kappa_m & 0 & 0 & 0\\
			-g_a & 0 & 0 & -\kappa_m & -G_m & 0\\
			0 & 0 & 0 & 0 & 0 & \omega_b\\
			0 & 0 & -G_m & 0 & -\omega_b+\mathrm{Re} \zeta(\omega) & -\gamma_b-\frac{\omega_b}{\omega}\mathrm{Im} \zeta(\omega)
		\end{pmatrix},
	\end{equation}
 where 
 \begin{equation}
 \zeta(\omega)=\sqrt{\eta}g_a G_m \chi_{a}(\omega) \chi_{ma}(\omega)  g(\omega). 
 \end{equation}
 From Eq.~\eqref{AAA}, we see that the feedback force modulates not only the mechanical dissipation rate, but also the mechanical resonance frequency. However, as will be shown later, the feedback gain function $g(\omega)$ can be properly chosen, such that the effective mechanical damping rate is significantly enhanced while the mechanical frequency remains unchanged.  

	To characterize the cooling effect of the mechanical mode, we look at the NSD of the mechanical position and momentum, which are defined as
	\begin{equation}
			S_q(\omega)=\frac{1}{4\pi}\int_{-\infty}^{+\infty} \langle \delta q(\omega)\delta q(\omega')+\delta q(\omega')\delta q(\omega) \rangle \ e^{i(\omega+\omega')t} {\rm d}\omega', 
	\end{equation}
and $S_p(\omega)=\frac{\omega^2}{\omega_b^2} S_q(\omega)$. By completely solving the QLEs related to the drift matrix $A$ in Eq.~\eqref{AAA}, the NSD of the mechanical position in such a feedback-assisted CMM system is given by 
	\begin{align}\label{Sqqq}
		S_{q}(\omega)=\left| \chi_b^{\mathrm{eff}}(\omega) \right|^2 \left[ S_{a}^{\rm ba}(\omega)+S_m^{\rm ba}(\omega)+S_b^{\rm th}(\omega)+S_{\mathrm{fb}}(\omega) \right],
	\end{align}
where $\chi_b^{\mathrm{eff}}(\omega)$ is the effective mechanical susceptibility, 
\begin{equation}
		\chi_b^{\mathrm{eff}}(\omega)= \frac{\omega_b}{\omega_b^2-\omega^2-i\gamma_b\omega-\zeta(\omega)\omega_b},
	\end{equation}
and 
	\begin{equation}
		\begin{split}
			S_{a}^{\rm ba}(\omega)=&\ g_a^2G_m^2\left | \chi_a(\omega) \chi_{ma}(\omega) \right |^2 \kappa_a(2n_a+1),  \\
			S_m^{\rm ba}(\omega)=&\ G_m^2\left | \chi_{ma}(\omega) \right |^2 \kappa_m(2n_m+1)
		\end{split}
	\end{equation}
denote the back-action noises from the cavity mode and the magnon mode, respectively; 
	\begin{equation}
		S_b^{\rm th}(\omega)= \frac{\gamma_b\omega}{\omega_b}\mathrm{coth}\left(\frac{\hbar\omega}{2k_B T}\right)
	\end{equation}
is the noise from the local thermal bath of the mechanical resonator; and $S_{\mathrm{fb}}(\omega)= S_{\mathrm{fb}}^{\mathrm{a,m}}(\omega)+S_q^{\mathrm{imp}}(\omega)$ represents all the noises introduced by the feedback loop, which can be separated into two noise sources: 
\begin{equation}\label{fbba}
		\begin{split}
			S_{\mathrm{fb}}^{\mathrm{a,m}}(\omega)=&\ \frac{\left | g(\omega) \right |^2 }{4\kappa_a}[\eta(2n_a+1)+(1-\eta)]\\
			&+\eta\left | g(\omega) \right |^2 g_a^2\left | \chi_a(\omega)\chi_{ma}(\omega) \right |^2\kappa_m(2n_m+1)\\
			&-\eta\left | g(\omega) \right |^2 g_a^2\left | \chi_a(\omega)\chi_{ma}(\omega) \right |^2 \kappa_m(2n_a+1)
		\end{split}
	\end{equation}
is the back-action noise from the cavity and magnon modes {\it through the feedback loop}, and 
	\begin{equation}\label{fbmea}
			S_{q}^{\mathrm{imp}}(\omega)= \left|g(\omega) \right|^2 S_{X_a}^{\mathrm{imp}}
	\end{equation}
is the feedback-gained measurement noise in the detection of the output field quadarature, with $S_{X_a}^{\mathrm{imp}}$ being the NSD of $X_a^{\mathrm{imp}}$.  It is worth noting that the back-action noise through the feedback loop $S_{\mathrm{fb}}^{\mathrm{a,m}}$ does not contain the noise from the local thermal bath of the mechanical resonator. This is the result of the choice of the resonant drive and the detection of the microwave amplitude quadrature. Under these conditions, the mechanical thermal noise affects only the cavity phase quadrature via their coupling to the magnon amplitude quadrature, as can be seen from Eqs.~\eqref{bbb}, \eqref{ccc}, and \eqref{fff}. Therefore, the mechanical thermal noise does not enter the feedback loop.

The feedback force applied onto the mechanical resonator modifies the mechanical susceptibility $\chi_b(\omega)= \omega_b/(\omega_b^2-\omega^2-i\gamma_b\omega)$ to be an effective one $\chi_b^{\mathrm{eff}}(\omega)$, and the corresponding changes of the mechanical frequency and damping rate are 
	\begin{equation}
		\begin{split}
			\delta\omega_b=&-\mathrm{Re}\zeta(\omega),\\
			\delta\gamma_b=&\ \frac{\omega_b}{\omega}\mathrm{Im}\zeta(\omega),
		\end{split}
	\end{equation}
which depend on the gain function $g(\omega)$. By utilizing the controllable gain module, e.g., the FPGA-based feedback circuit, we can delicately design the gain function to take the following form:
	\begin{equation}\label{ggg}
		g(\omega)=\frac{i\gamma_b\omega g_0}{\sqrt{\eta}\omega_b g_a G_m\chi_a(\omega )\chi_{ma}(\omega)},
	\end{equation}
with $g_0$ being a dimensionless gain coefficient. In practice, the feedback gain is applied near the mechanical resonance and in a proper bandwidth. Specifically, we apply the feedback gain in the frequency range $\omega\in \{-2\omega_b,2\omega_b\}$ by sending the output field through a filter (with the central frequency $\omega=0$ and bandwidth of $4 \omega_b$).  The feedback bandwidth should be appropriately chosen: on the one hand, the feedback cannot give full play to the cooling effect when the bandwidth is too narrow; on the other hand, a large amount of noise will be added when the bandwidth is too wide~\cite{c3}.  The choice of the gain function in Eq.~\eqref{ggg} leads to a pure imaginary $\zeta(\omega)$ and thus a zero mechanical frequency shift $\delta\omega_b =0$ and an increased mechanical damping rate 
\begin{equation}\label{gammaeff}
\gamma_b^{\mathrm{eff}} \equiv \gamma_b +\delta \gamma_b = (1+g_0)\gamma_b, 
\end{equation}
for $g_0 \gg 1$. This reflects the advantages of the measurement-based feedback cooling, i.e., the feedback force can significantly enhance the mechanical damping rate but meanwhile keep the mechanical frequency unaffected. This also results in the drift matrix $A$ in the form of
	\begin{equation}
		A=\begin{pmatrix}
		-\kappa_a & 0 & 0 & g_a & 0 & 0\\
		0 & -\kappa_a & -g_a & 0 & 0 & 0\\
		0 & g_a & -\kappa_m & 0 & 0 & 0\\
		-g_a & 0 & 0 & -\kappa_m & -G_m & 0\\
		0 & 0 & 0 & 0 & 0 & \omega_b\\
		0 & 0 & -G_m & 0 & -\omega_b & -\gamma_b^{\mathrm{eff}}
		\end{pmatrix}.
	\end{equation}
The system reaches a steady state when $t \to \infty$ if all the eigenvalues of the drift matrix $A$ have negative real parts. This is checked to guarantee that all the results presented in Sec.~\ref{result} are in the steady state.
		
	Finally, we calculate the effective mean phonon number to investigate the ground-state cooling of the mechanical mode, which can be achieved by 
	\begin{equation}
		\bar{n}_b^{\mathrm{eff}}=\frac{1}{2}\left(\langle \delta q(t)^2\rangle+\langle \delta p(t)^2\rangle-1\right),
	\end{equation}
	where 
	\begin{equation}
	\begin{split}
 \langle \delta q(t)^2\rangle&=\frac{1}{2\pi}\int_{-\infty}^{+\infty}S_q(\omega)\ {\rm d}\omega, \\
  \langle \delta p(t)^2\rangle&=\frac{1}{2\pi}\int_{-\infty}^{+\infty}S_p(\omega)\ {\rm d}\omega 
        \end{split}
	\end{equation}
are the variance of the mechanical position and momentum, respectively.  The mechanical mode is cooled into its ground state if $\bar{n}_b^{\mathrm{eff}}<1$.

	\section{Results}\label{result}


	\begin{figure}[b]
		\includegraphics[width=0.85\linewidth]{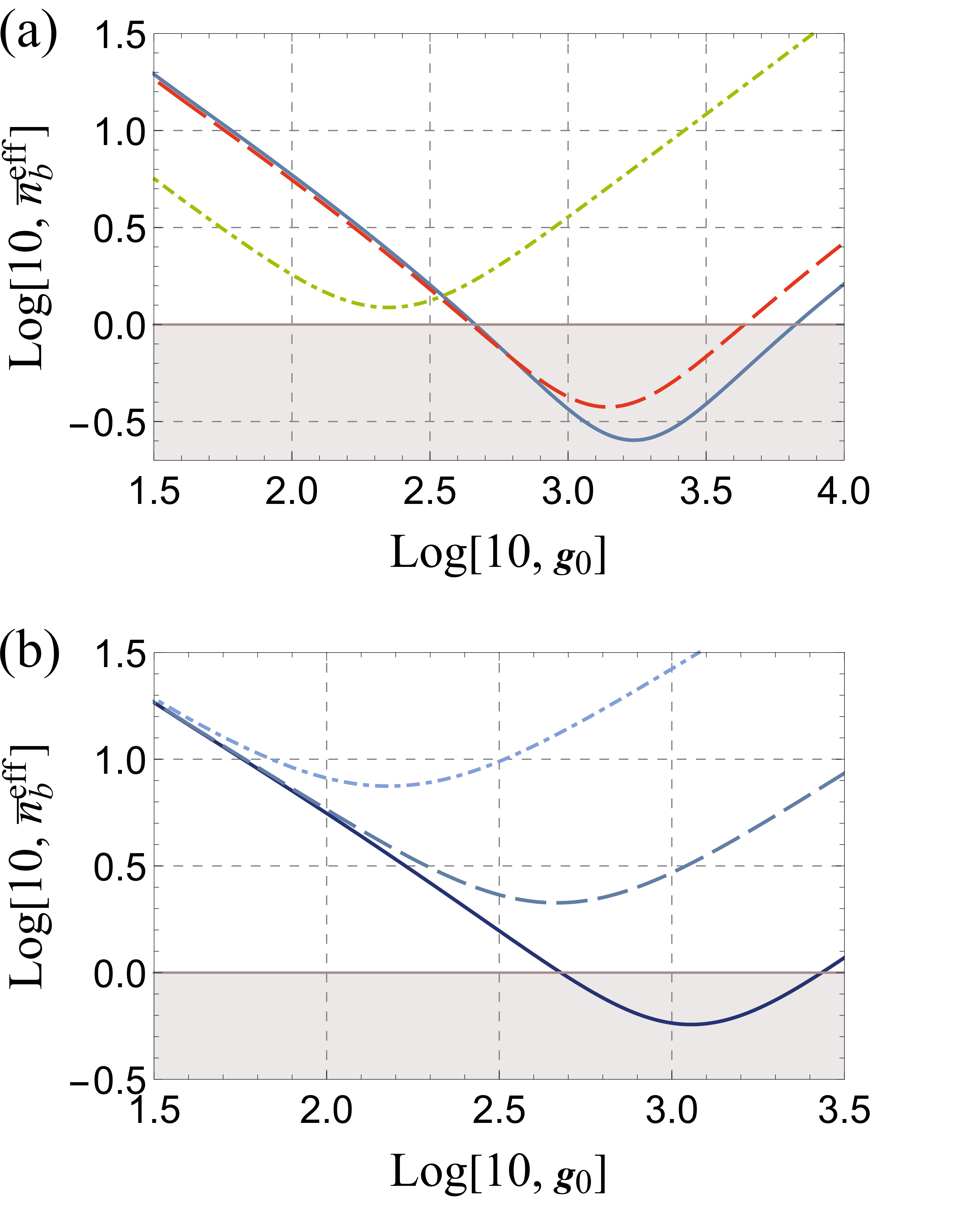}
		\caption{Steady-state effective mean phonon number $\bar{n}_b^{\mathrm{eff}}$ versus gain coefficient $g_0$ (a) for $\kappa_m/2\pi=1$ MHz (solid), $10$ MHz (dashed), and $10^2$ MHz (dot-dashed); (b) for $S_{X_a}^{\mathrm{imp}}=10^{-8}$ Hz$^{-1}$ (solid), $10^{-7}$ Hz$^{-1}$ (dashed), and $10^{-6}$ Hz$^{-1}$ (dot-dashed). We take $S_{X_a}^{\mathrm{imp}}=6.65 \times10^{-9}$ Hz$^{-1}$, $4.04 \times 10^{-9}$ Hz$^{-1}$ and $8.22 \times 10^{-10}$ Hz$^{-1}$ for the solid, dashed and dot-dashed lines in (a), respectively, and $\kappa_m/2\pi=10$ MHz in (b). See text for the other parameters.}
		\label{fig2} 
	\end{figure}

	\begin{figure}[t]
		\includegraphics[width=0.85\linewidth]{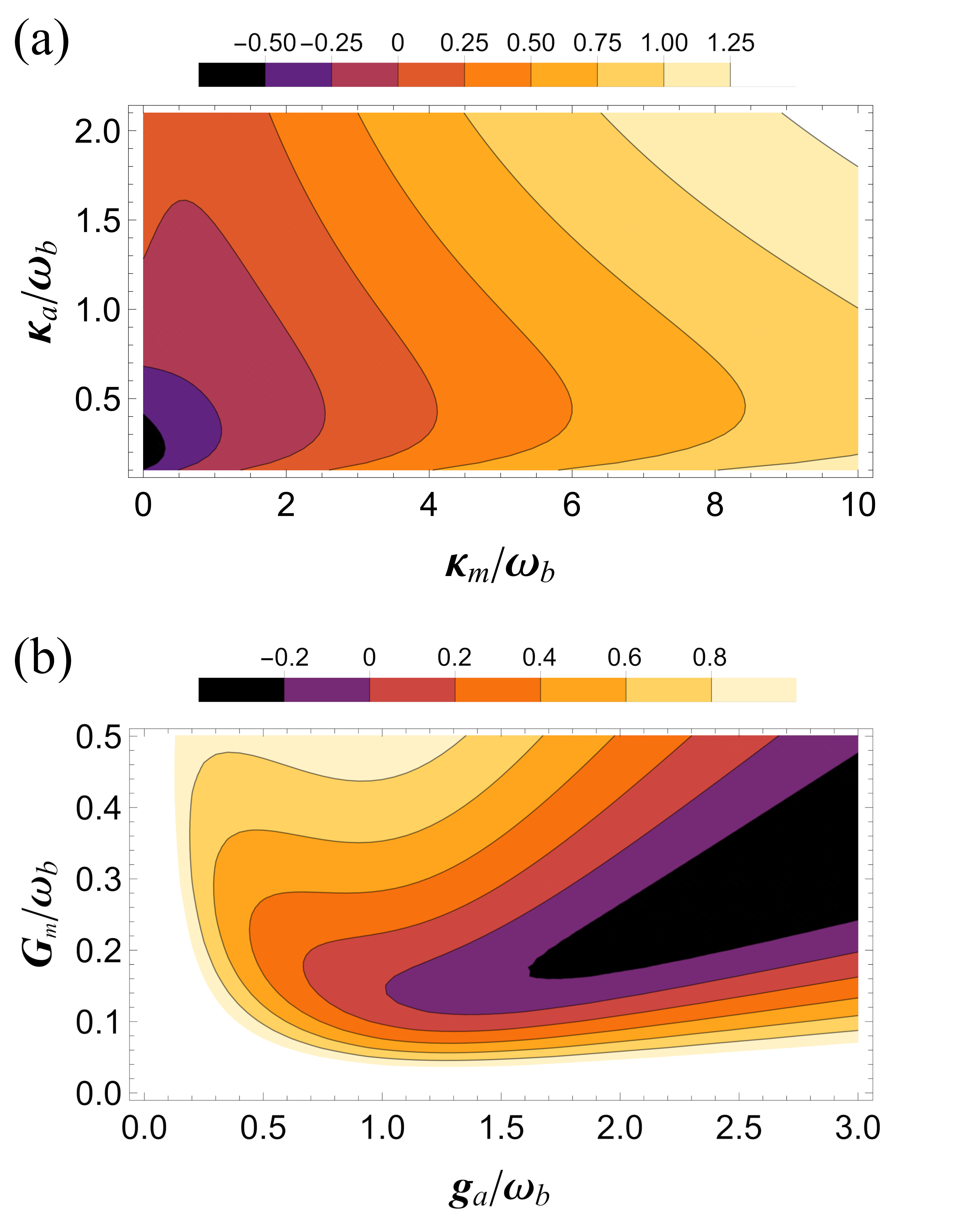}
		\caption{Contour plot of ${\rm Log}[10, \bar{n}_b^{\mathrm{eff}}]$ versus (a) dissipation rates $\kappa_a$ and $\kappa_m$; (b) cavity-magnon coupling strength $g_a$ and magnomechanical coupling strength $G_m$, with the gain coefficient $g_0=10^3$. We take $\kappa_m/2\pi=10\ \mathrm{MHz}$ in (b), and $S_{X_a}^{\mathrm{imp}}=10^{-8}$ Hz$^{-1}$ for both plots.  The blank area in (b) denotes $\bar{n}_b^{\mathrm{eff}} >10$. The other parameters are the same as in Fig.~\ref{fig2}(a).}
		\label{fig3}
	\end{figure}

In this section, we present the main results of the feedback-assisted mechanical cooling.  Figure~\ref{fig2}(a) shows the steady-state effective mean phonon number $\bar{n}_b^{\rm eff}$ versus the gain coefficient $g_0$ for three values of the magnon dissipation rate, $\kappa_m/2\pi=1$ MHz, $10$ MHz, and $10^2$ MHz. We employ the following experimentally feasible parameters: $\omega_a/2\pi\simeq\omega_m/2\pi=10\ \mathrm{GHz}$, $\omega_b/2\pi=10\ \mathrm{MHz}$, $\gamma_b/2\pi=10^2\ \mathrm{Hz}$, $\kappa_a/2\pi=5\ \mathrm{MHz}$, $\eta=0.9$, $g_a/2\pi=18\ \mathrm{MHz}$, $G_m/2\pi=2\ \mathrm{MHz}$, and $T=10\ \mathrm{mK}$, which corresponds to the mean thermal phonon number $\bar{n}_b \simeq 20$. 
 Figure~\ref{fig2}(a) reveals that by properly choosing the gain coefficient $g_0$, the mechanical mode can be cooled into its quantum ground state not only in the resolved-sideband limit $\kappa_m/2\pi=1$ MHz $\ll \omega_b$ with a minimum $\bar{n}_b^{\mathrm{eff}} \simeq 0.25$, but also in the unresolved-sideband regime $\kappa_m =\omega_b$ with a minimum $\bar{n}_b^{\mathrm{eff}} \simeq 0.38$. For a significantly large dissipation rate $\kappa_m = 10\omega_b$, our feedback cooling protocol is still efficient and the mechanical mode is cooled from the mean occupation of $\bar{n}_b \simeq 20$ to $\bar{n}_b^{\mathrm{eff}}\simeq 1.22$ at an optimal $g_0$.  The presence of an optimal gain $g_0$ is a result of the trade-off between the feedback-induced significant enhancement of the mechanical damping rate (Eq.~\eqref{gammaeff}) and the additional noises (Eqs.~\eqref{fbba} and \eqref{fbmea}) introduced by the feedback loop.     Note that the measurement noise $S_{X_a}^{\mathrm{imp}}$ can be estimated by using the feedback cooling and quantum measurement theories~\cite{15fbNature,RMPmeasure}. We obtain $S_{X_a}^{\mathrm{imp}} \ge$ $6.65\times10^{-9}$ Hz$^{-1}$, $4.04\times10^{-9}$ Hz$^{-1}$ and $8.22\times10^{-10}$ Hz$^{-1}$ for $\kappa_m/2\pi=1$ MHz, $10$ MHz and $10^2$ MHz, respectively, under the parameters of Fig.~\ref{fig2} using the uncertainty relation between the back-action noise and the measurement noise~\cite{15fbNature}.

In Fig.~\ref{fig2}(b), we show the results in the unresolved-sideband regime ($\kappa_m =\omega_b$) for three values of $S_{X_a}^{\mathrm{imp}}=10^{-8}$, $10^{-7}$, and $10^{-6}$ Hz$^{-1}$.  Correspondingly, we obtain the effective mean phonon number $\bar{n}_b^{\mathrm{eff}} \simeq 0.57$, $2.12$, and $7.5$, respectively.    Clearly, the measurement noise should be kept to a minimum level, such that the feedback-gained measurement noise will not appreciably heat the mechanical mode.

 We further explore the impact of other key parameters of the system on the mechanical cooling. Specifically, we show $\bar{n}_b^{\mathrm{eff}}$ versus the dissipation rates $\kappa_m$ and $\kappa_a$ in Fig.~\ref{fig3}(a); and the coupling strengths $g_a$ and $G_m$ in Fig.~\ref{fig3}(b).  Figure~\ref{fig3}(a) clearly reveals that the feedback cooling is efficient in the unresolved-sideband regime: the mechanical mode can be cooled into its ground state (with $\bar{n}_b^{\mathrm{eff}} <1$) for $\kappa_m$ up to $\sim2\omega_b$, and can be cooled (with $\bar{n}_b^{\mathrm{eff}} < \bar{n}_b$) even for a very large $\kappa_m \gg \omega_b$. For a specific ferromagnetic material with a certain magnon dissipation rate, there is an optimal cavity decay rate to achieve the minimum $\bar{n}_b^{\mathrm{eff}}$.
Figure~\ref{fig3}(b) shows that the mechanical ground state can be reached when $g_a > \kappa_m, \kappa_a$, and there is an optimal magnomechanical coupling $G_m$ for a given cavity-magnon coupling $g_a$. In general, a large $g_a$ is helpful to achieve the minimum of $\bar{n}_b^{\mathrm{eff}}$. The existence of an optimal value of the parameter is the result of the different dependences of the noise terms in $S_{q}(\omega)$ (Eq.~\eqref{Sqqq}) on the parameter.

	\begin{figure}[t]
	\includegraphics[width=0.75\linewidth]{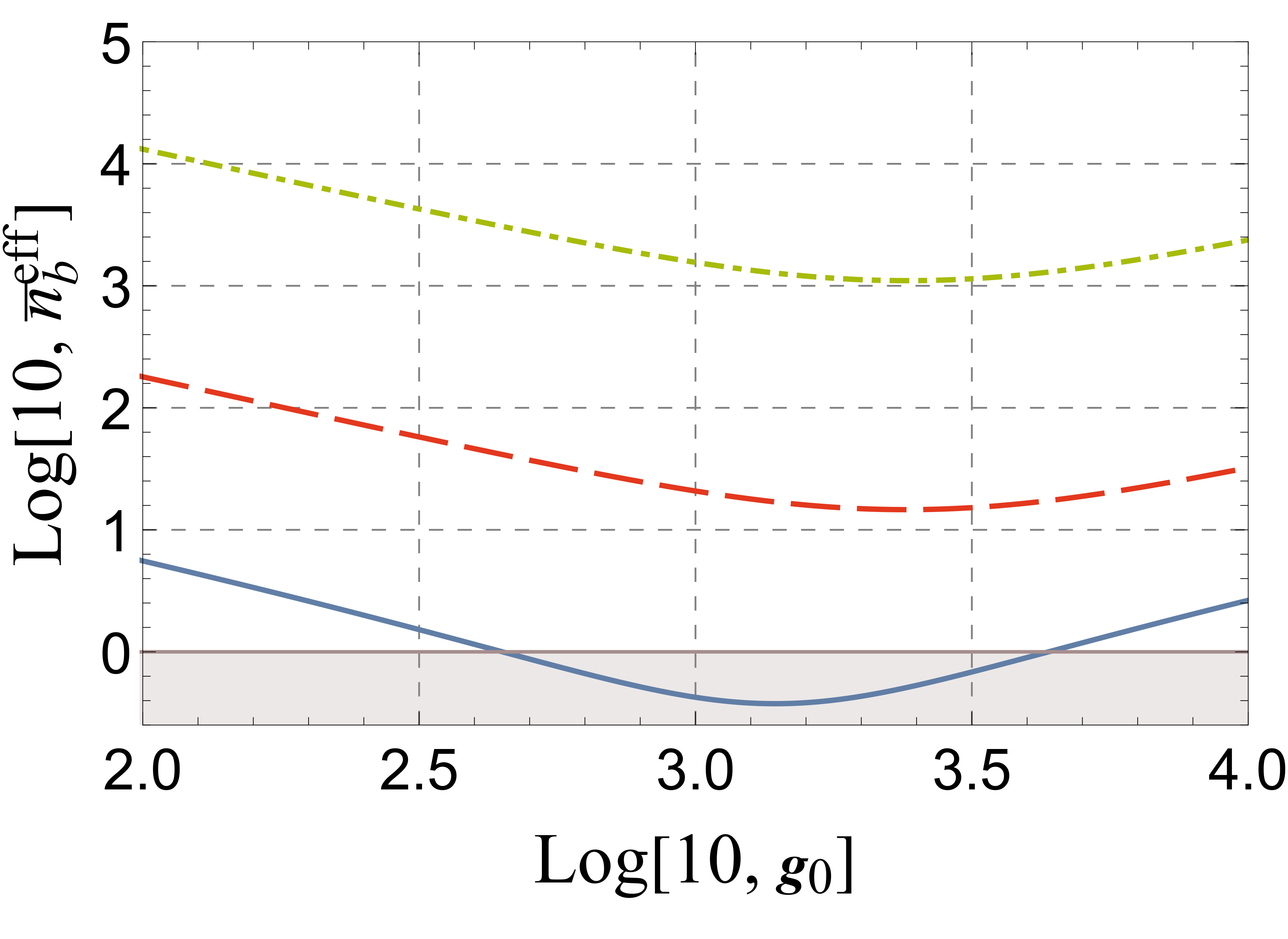}
		\caption{Effective mean phonon number $\bar{n}_b^{\mathrm{eff}}$ versus gain coefficient $g_0$ at $T= 10$ mK (solid), 4 K (dashed), and 293 K (dot-dashed), respectively. We take $\kappa_m/2\pi=10$ MHz, and $S_{X_a}^{\mathrm{imp}}=4.04\times10^{-9}$ Hz$^{-1}$, $2.42\times10^{-10}$ Hz$^{-1}$, $3.31\times10^{-12}$ Hz$^{-1}$ for $T= 10$ mK, 4 K, and 293 K,  respectively.  The other parameters are same as in Fig.~\ref{fig2}(a).}
		\label{fig4}
	\end{figure}

Lastly, we check our cooling scheme at higher bath temperatures. In Fig.~\ref{fig4}, we show $\bar{n}_b^{\mathrm{eff}}$ at bath temperatures $T= 10$ mK, 4 K, and 293 K, respectively. The mechanical mode is cooled down to $\bar{n}_b^{\mathrm{eff}}\simeq 0.38$, $15$, and $1.1\times 10^3$, respectively, from the thermal occupation $\bar{n}_b\simeq 20$, $8.3 \times 10^3$, and $6.1\times 10^5$ (corresponding to $T=10$ mK, 4 K, and 293 K). In general, the mechanical mode can be significantly cooled in the unresolved-sideband regime over a wide range of temperatures. Note that at each temperature, we reevaluate the minimum $S_{X_a}^{\mathrm{imp}}$ as the back-action noise increases as the temperature rises.

\section{Conclusion}\label{conc}
	
We have proposed a measurement-based feedback cooling protocol designed for CMM systems in the unresolved-sideband regime, where the magnon dissipation rate is comparable to or even much larger than the mechanical resonance frequency. By appropriately designing the feedback gain, the mechanical dissipation rate can be significantly enhanced, which allows for cooling the mechanical motion into its quantum ground state. Our feedback cooling, together with the sideband cooling~\cite{Jie18,Jie19b,Ding20}, form a complete theory for realizing mechanical cooling in the CMM systems with the magnon dissipate rate either $\kappa_m < \omega_b$ (sideband cooling) or $\kappa_m \ge \omega_b$ (feedback cooling). The work offers the possibility of observing macroscopic quantum states in the CMM systems based on ferromagnetic materials which show strong magnetostriction but also large magnon dissipation.  It may also find potential applications in quantum memory and high-precision measurement.

\section*{ACKNOWLEDGMENTS}

This work has been supported by National Key Research and Development Program of China (Grant No. 2022YFA1405200) and National Natural Science Foundation of China (Nos. 92265202 and 11874249).

\end{document}